\documentclass[%
 reprint, 
 amsmath,amssymb,
 aps,
]{revtex4-2}

\usepackage{float}
\usepackage{comment}
\usepackage{tabularx}
\usepackage{soul}

\usepackage[utf8]{inputenc}
\usepackage[T2A,T1]{fontenc}
\usepackage[english]{babel}

\usepackage{color}
\usepackage{ulem}
\usepackage{xcolor}
\usepackage{comment}

\usepackage{xfrac}

\usepackage{gensymb}

\usepackage[caption=false]{subfig}

\usepackage[toc]{appendix}

\usepackage{graphicx}
\usepackage{dcolumn}
\usepackage{bm}

\begin{document}


\title{Twist-tunable moiré optical resonances}

\author{Natalia S. Salakhova}
\email{Natalia.Salakhova@skoltech.ru}
\affiliation{Skolkovo Institute of Science and Technology, Bolshoy Boulevard 30, bld. 1, Moscow 121205, Russia}
\author{Ilia M. Fradkin}
\affiliation{Skolkovo Institute of Science and Technology, Bolshoy Boulevard 30, bld. 1, Moscow 121205, Russia}
\affiliation{Moscow Institute of Physics and Technology, Institutskiy pereulok 9, Moscow Region 141701, Russia}
\author{Sergey A. Dyakov}
\affiliation{Skolkovo Institute of Science and Technology, Bolshoy Boulevard 30, bId. 1, Moscow 121205, Russia}
\author{Nikolay A. Gippius}
\affiliation{Skolkovo Institute of Science and Technology, Bolshoy Boulevard 30, bId. 1, Moscow 121205, Russia}
\date{\today}

\begin{abstract}

Multilayer stacks of twisted optical metasurfaces are considered as a prospective platform for chiral nanophotonic devices. Such structures are primarily used for the realization of circularly polarized light sources, artificial optical rotation, and circular dichroism. At the same time, the behavior of their hybrid photonic modes is strongly affected by the moiré-pattern of superimposed periodic constituents. In this work, we show that moiré-periodicity in bilayer dielectric photonic crystal slabs leads to an arise of unlimitedly narrow optical resonances, which are very sensitive to the relative twist and gap width between the sublayers. We demonstrate the structure providing twist-tuning of the hybrid mode wavelength in the range of 300--600~nm with quality factor varying from~$10^2$~up~to~$10^5$ correspondingly. The obtained results pave the wave for the utilization of moiré-assisted effects in multilayer photonic crystal slabs.
\end{abstract}

\maketitle


Optical metasurfaces and photonic crystal slabs represent a notable branch of modern nanophotonics. They underlie a plethora of devices designed over the past several decades for the purposes of sensing~\cite{Chang2018,Wang2020,lodewijks2012,rodriguez2011,Shen2013}, resonant field enhancement~\cite{Yuan2017a,Krasnok2016,Yang2018}, second or third harmonics generation~\cite{Liu2019a,Smirnova2016,Shcherbakov2014,Wang2018}, nanolasing~\cite{Noda2017,wang2017,melentiev2017plasmonic,zyablovsky2017optimum}, on-chip light manipulation~\cite{fradkin2020nanoparticle,fradkin2022plasmonic}, holography~\cite{zheng2015metasurface,ye2016spin,wei2017broadband,ignatov2018excitation}, flat optics~\cite{yu2014flat,chen2016review,kildishev2013planar}, etc. One of the ways to enrich the available options of the already diversified world of periodic optical structures is to introduce the hybridization of their modes. This concept includes lattices with complex unit cells~\cite{baur2018,kolkowski2019lattice,Humphrey2014, zundel2021lattice}, plasmon-photonic structures~\cite{Linden2001,fradkin2022plasmonic}, stacks of metasurfaces~\cite{fradkin2020thickness,voronin2022single,becerril2021optical,murai2021photoluminescence,chen2019all,chen2019empowered,berkhout2019perfect} and many other examples.
In an attempt to find new ways to engineer structures with desired optical properties, the famous concept of twistronics~\cite{Balents2020,Chen2019,Cao2018,Pixley2019,Sharpe2019,Chen2020,Repellin2020,Li2010,Jiang2019charge,Slagle2020,Andrei2021} was applied in purely optical devices as well.
Twisted stacks are primarily known as chiral structures for efficient interaction with circularly polarized light~\cite{Ma2018,Xu2019,Zhao2017,Hu2021,aftenieva2021tunable,Salakhova2021}, but their potential is much wider.
In particular, recent studies~\cite{Hu2020,Hu2020a,Chen2020moire,Duan2020,Zheng2020,Salakhova2021,nikitin2022twist,wang2020localization,torrent2020dipolar,aftenieva2021tunable,zeng2021localization} demonstrated rich opportunities for dispersion control in twisted bilayers of 2D materials, light localization, so-called field canalization and topological transitions between closed and open isofrequency contours. 
Moreover, the relative rotation of periodic structures leads to an appearance of a superlattice~\cite{Salakhova2021}, which is associated with widely known moiré patterns. As we demonstrated in our previous paper~\cite{Salakhova2021}, the interaction between the lattices results in the excitation of not only the diffraction harmonics of each sublattice but also the hybrid, moiré-induced ones. In turn, these harmonics lead to the arise of some peculiar moiré modes that require detailed analysis.

\begin{figure}[ht]
    \centering
    \includegraphics[width=0.9\linewidth]{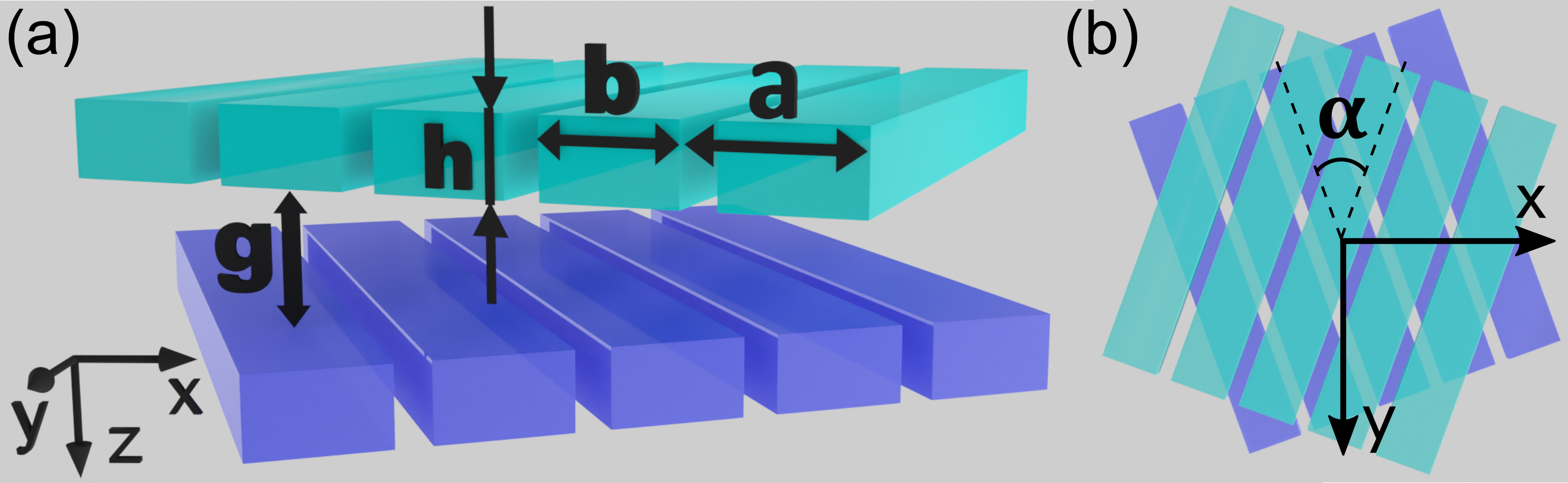}
    \caption{The sketch of the moiré structure comprised twisted photonic crystal slabs: (a) general view and (b) top view.}
    \label{fig:1}
\end{figure}

Here, we study the interaction of two twisted photonic crystal slabs sketched in Fig.~\ref{fig:1}. Being equipped with our domestically-developed Moiré-Adapted Fourier modal method (MA-FMM)~\cite{Salakhova2021}, we calculate spectral maps of the twisted structures and demonstrate extremely narrow mechanically-tunable moiré-induced modes. We show that these hybrid modes are highly sensitive to the rotation of sublayers and the distance between them. In particular, we achieve a twofold spectral shift of the mode wavelength (between 300~nm and 600~nm) and quality factor variation in the range from $10^2$ to $10^5$. Our results pave the way for the practical engineering of twisted metasurfaces with high-quality, tunable optical modes for modern nanophotonics.




\begin{figure*}
    \centering
    \includegraphics[width=0.65\textwidth]{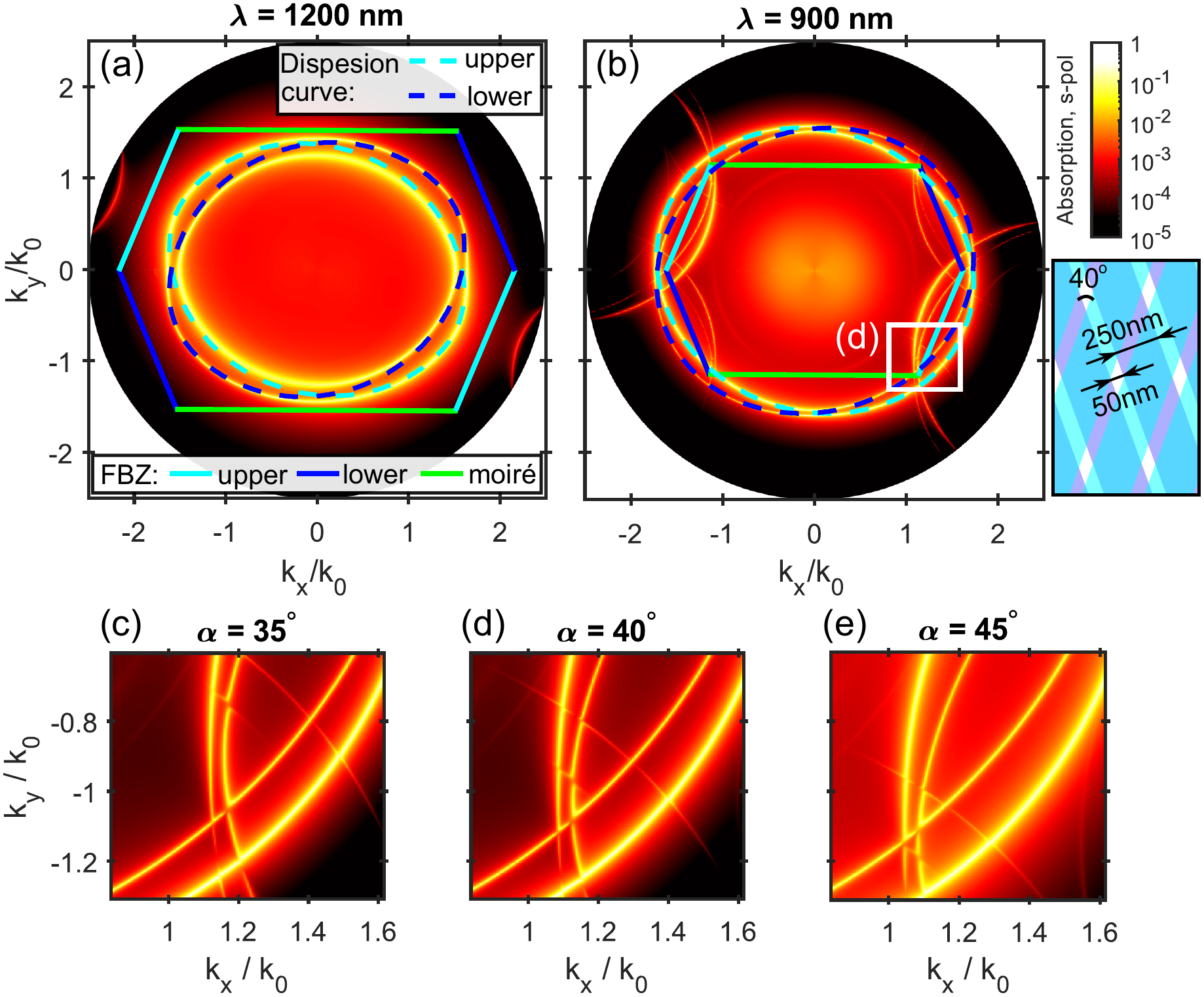}
    \caption{Absorption maps of the stack of two identical 1D PCSs  twisted by $\alpha=40^\circ$ for the $s$-polarized incident light with the wavelength (a) $\lambda=1200$~nm and (b) $\lambda=900$~nm. The dashed cyan (blue) lines show the quasiguided modes of the upper (lower) photonic crystal slab in absence of the lower (upper) one. Solid lines of cyan and blue colors in panels (a) and (b) denote boundaries of the FBZ corresponding to the periodicity of the upper and lower lattices. The green lines stand for the moiré-induced boundaries of the FBZ. Panel (d) shows the enlarged version of the area indicated by the white box in panel (b), whereas panels (c,e) demonstrate maps corresponding to the lattices of slightly different rotation angles $\alpha=40^{\degree}\pm 5 ^\degree$.}
    \label{fig:2}
\end{figure*}

In this study, we consider the structure of two identical lamellar photonic crystal slabs. Figure~\ref{fig:1} shows the vacuum-surrounded lattices made of a relatively high-index dielectric. They have a rectangular profile, the period of each sublayer is $a$, the width of dielectric strips is $b$ and the height is $h$. Upper and lower photonic crystal slabs, separated by the gap, $g$, are rotated in opposite directions each by $\alpha/2$. As a result, we deal with a moiré-periodic structure possessing second-order rotation symmetry for the x- and y-axes.
In the scope of our study, it is important that each of the sublayers supports photonic guided modes by itself. As a result of their interaction in specific moiré geometry, we obtain hybrid modes of unique optical properties.

The formation of a 2D-periodic moiré lattice makes the structure much more complicated for numerical consideration in contrast with its 1D-periodic constituents. In particular, the small value of rotation angle, $\alpha$, makes the unit cell of a superlattice correspondingly large and strongly increases the number of opened diffraction channels. As a result, both real-space-based numerical approaches such as finite element method (FEM), finite difference time domain (FDTD), and reciprocal-space based Fourier modal method (FMM)~\cite{Tikhodeev2002,moharam1995} become inefficient. For this reason, we apply our domestic-developed Moiré-Adapted FMM (MA-FMM)~\cite{Salakhova2021}, which accounts for the peculiar structure of the considered 2D photonic crystal to obtain the scattering matrix of the twisted stack in a reasonable time.



To study the impact of the moiré pattern on the quasiguided modes of the double-slab system, we calculate the in-plane wavevector dependence of the absorption coefficient of incident light. The results are obtained for the following structure parameters: the gratings' period is $a=300$~nm, the dielectric strips' width is $b=250$~nm, the thickness of each slab is $h=150$~nm,  the gap size is $g=300$~nm, and rotation angle $\alpha=40\degree$. The dielectric constant of the gratings material is $\varepsilon = 6.25+0.01i$. We add a small imaginary part to the dielectric permittivity to enable the absorption spectra that clearly indicate the excitation of resonant modes (see Fig.~\ref{fig:2}).

\begin{figure*}[ht]
    \centering
    \includegraphics[width=0.9\textwidth]{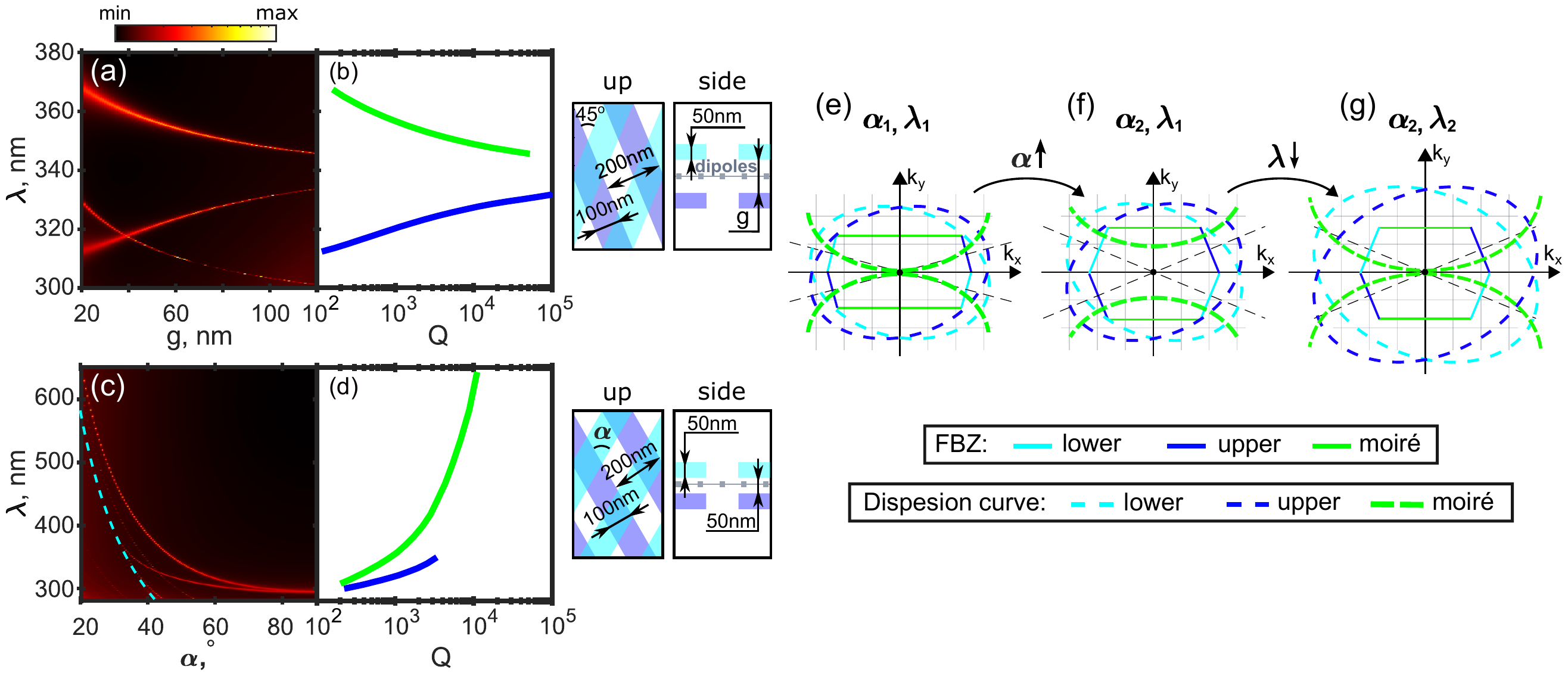}
    \caption{The emissivity of uniformly distributed oscillating dipoles placed in the middle of the gap layer  (a) as a function of the wavelength and the thickness of the gap layer and (c) as a function of the wavelength and the total rotation angle. In panel (c), the dashed cyan line denotes the opening of the lowest, moiré-induced diffraction channel.    Panels (b) and (d) show the quality factors $Q$ of the corresponding modes from panels (a) and (c). Schematics of the considered structures are sketched next to panels (b) and (d).    Panels (e)-(g) show the shift of a moiré mode under the change of the relative rotation angle and wavelength of light. Boundaries of the first Brillouin zone are denoted by solid lines, while the modes' isofrequency contours are shown by dashed lines. The cyan (blue)    lines correspond to the modes and boundaries of the first Brillouin zone corresponding to the upper (lower) sublattice,    while the green lines correspond to the moiré-induced modes and boundaries.}
    \label{fig:3}
\end{figure*}

\begin{figure*}[ht]
    \centering
    \includegraphics[width=0.85\linewidth]{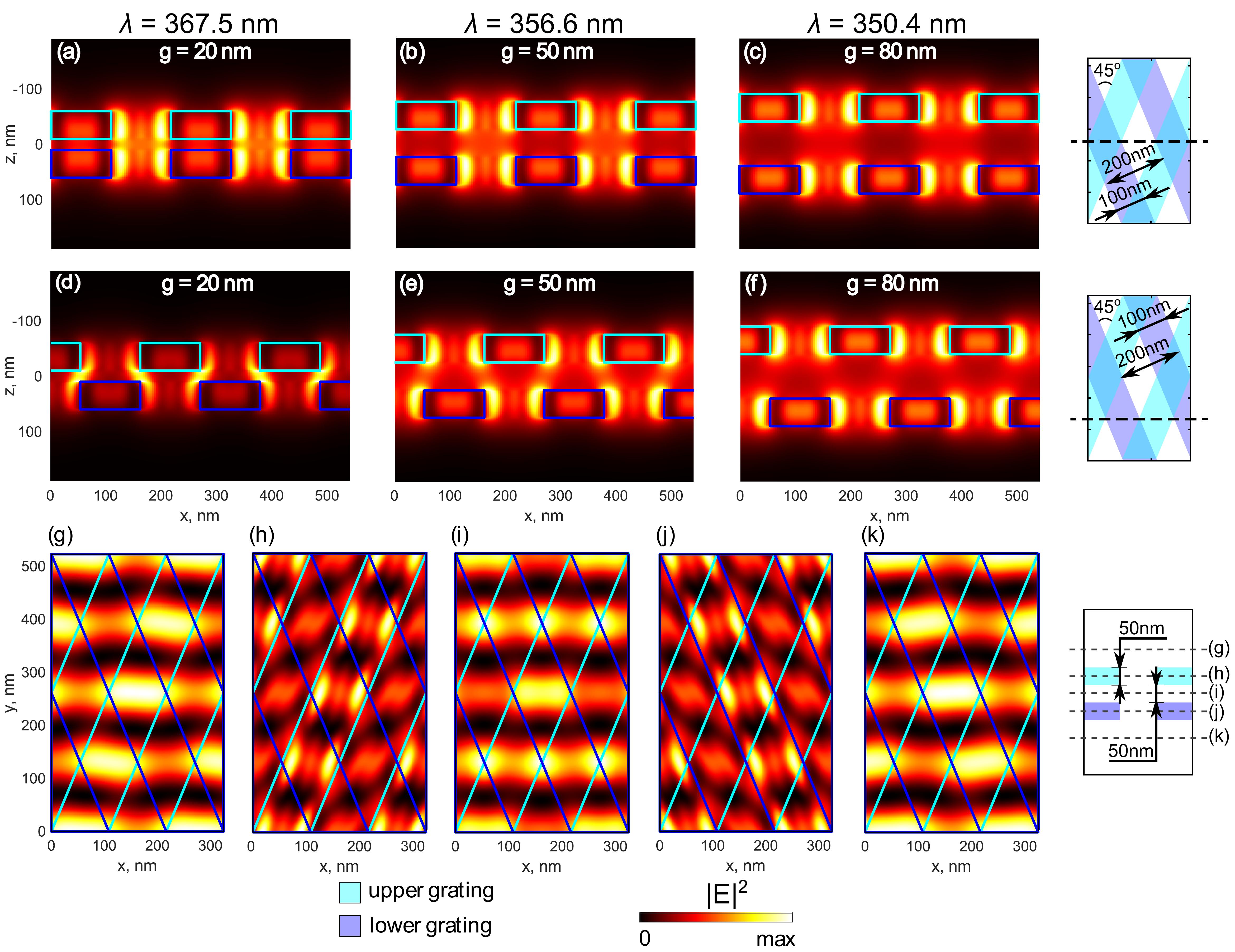}
    \caption{The eigenmode electric field intensity in different sections for structure with parameters $a=200$~nm, $b=100$~nm, $h=50$~nm, and $\alpha=45^\degree$. Panels (a,b,c) and (d,e,f) show the field distribution in $x-z$ (vertical) plane for three different thicknesses of the gap layer $g = 20$~nm, $50$~nm  and $80$~nm and for two different cross-sections indicated in the right column.  Panels (g)-(k) show the electric field distribution in five $x-y$ (horizontal) planes  The uppermost (panel (g)) and lowermost (panel (k)) horizontal cross-section are taken at $50$~nm distance from the nearest interface, while the rest of the horizontal cross-sections (panel (h-j)) lay in the middle of the layers. The dielectric stripes of the upper (lower) layer are denoted by cyan (blue) color.}
    \label{fig:4}
\end{figure*}

We start with the study of the quasiguided modes' isofrequency dispersion. To excite the quasiguided modes, we utilize the coupling prism with permittivity $\varepsilon_{\mathrm{prism}} = 6.25$ that is placed at a distance $H=300$~nm from the upper lattice. The absorption map for the $s$-polarized 1200-nm-wavelength light is shown in Fig.~\ref{fig:2}~(a). 
It is seen that the dispersion consists of two twisted ellipses, which correspond to the upper and lower sublayers. As a result of their interaction, we observe small areas of avoided crossing.
In our scheme, the structure is excited by the strongly attenuating evanescent waves, so the excitation of the mode depends on the distance from the prism. Indeed, the brighter mode rotated in a clockwise direction relates to the upper lattice. Whereas, the darker mode rotated in the counterclockwise direction corresponds to the lower lattice.

If we consider the same spectrum for a slightly smaller 900~nm wavelength of light (see Fig.~\ref{fig:2}~(b)), we immediately observe an arise of many additional branches of the guided modes. Indeed, they appear as a result of folding to the first Brillouin zone (FBZ) of the modes that expanded beyond its boundaries. It is worth noting that there are three pairs of FBZ boundaries. Two of them indicated by cyan and blue lines correspond to the diffraction on the upper and lower lattices respectively and are located at a distance $\pi/a$ from the $\Gamma$ point. The green pair corresponds to the moiré-induced diffraction along the $y$-axis, which gives us $k_y=\pm\pi\sin(\alpha/2)/a$ boundaries in reciprocal space. Let us take a closer look at the region near the right bottom vertex of the FBZ (see Fig.~\ref{fig:2}~(d) for an enlarged region from Fig.~\ref{fig:2}~(b)). It is clearly seen that some of the modes are "reflected" from the cyan, lamellar-grating-assisted boundaries, but the most narrow, barely noticeable modes originate as a "reflection" from the horizontal green boundary. In other words, they arise as a result of a coupling between the sublattices and therefore fully depend on their relative position and strength of the interaction. In particular, even a slight change of the rotation angle, $\alpha$, results in a shift of the green boundaries of the FBZ ($k_y=\pm\pi\sin(\alpha/2)/a$) and corresponding moiré modes. The absorption maps for three close angles $\alpha=35^\degree,40^\degree,45^\degree$ (see Fig.~\ref{fig:2}~(c-e)) demonstrate that $5^\degree$ deviation almost does not affect the wide modes of upper and lower sublattices, but strongly shift the narrow lines of hybrid resonances.

So far we used the coupling prism for the excitation of the system eigenmodes from the far field. Although such an approach is a rather simple way of plotting dispersion curves, it cannot be implemented for the characterization of the quality factor of the modes. Indeed, the exciting prism automatically enables the leakage of the modes. Moreover, some of the losses come from the material absorption in the slabs, which we introduced previously. Therefore, to study the quality factor of the moiré modes, from now on we use a non-dissipative slab ($\varepsilon=6.25+0i$) and excite the eigenmodes by uniformly distributed dipoles oscillating within the $xy$ plane; dipoles are placed in the middle of the gap layer and emit noncoherent light. Geometric parameters of the structure are chosen in such a way that the isofrequency contours of the moiré-induced modes pass through the $\Gamma$-point (see sketches in Fig.~\ref{fig:3} for the values of $a=200$~nm, $b=100$~nm, $h=50$~nm). To study the hybrid modes we calculate far-field emissivity spectra in the normal direction (see Ref.~\cite{lobanov2012emission, dyakov2020vertical} for details on the calculation of dipoles emission using the Fourier modal method). Our primary interest is concentrated on the dependence of the modes on the relative position of the sublattices, which is characterized by the gap size $g$ and angle $\alpha$.

The calculated gap-size dependence of the emissivity spectrum is shown in Fig.~\ref{fig:3}~(a) for the rotation angle $\alpha = 45^\degree$. It can be seen, that the smaller the gap size, the stronger the near-field interaction between the sublattices' eigenmodes and, hence, the larger the spectral distance between the hybridized moiré modes. When the gap size between two sublattices is relatively large the interaction between them weakens and lines of hybrid modes approach each other. Moreover, their quality factor rapidly increases and tends to infinity (see Fig.~\ref{fig:3}~(b)). Indeed, the existence of the hybrid mode at the $\Gamma$ point is determined by the coupling between the upper and lower sublattices, which is ensured by the evanescent, exponentially decaying harmonics. For a large distance between the constituents of the stack, coupling strength becomes a small parameter, whose non-zero value determines the finite width of the mode. 
Therefore, the variation of the gap size might be used to control the spectral position and the quality factor of the moiré-induced modes in a wide range. In the considered case, the gap size in the range of $20$ to $120$~nm provides the spectral shift of hybrid modes by up to $\approx20$~nm and simultaneously changes their quality factor within the range from 10$^2$ to 10$^5$ (see Fig.~\ref{fig:3}~(a-b)).

Let us now study how the rotation of the PCSs affects the hybrid modes. For this purpose, we calculate the dependence of the emissivity spectra on the relative rotation angle, $\alpha$, (Fig.~\ref{fig:3}~(c)). The cyan dashed line shows the first of the so-called Rayleigh anomalies, which in our case indicates the opening of the moiré-induced diffraction channel. Both hybrid modes of the structure are observed above the cyan line, where the diffraction is closed and leakage losses are suppressed. For the case of perpendicular sublattices ($\alpha = 90^{\degree}$) there are two mirror symmetry planes, which makes the modes degenerate. The spectral position and width of each resonance are very sensitive to the $\alpha$ angle (see Fig.~\ref{fig:3}~(c,d)).
In the presented structure with the gap size $g = 50$~nm, by changing the relative rotation angle from $\alpha=90^\degree$ to $\alpha=20^\circ$, we can tune the wavelength of the moiré-induced modes in a wide range from $300$~nm to $650$~nm (see Fig.~\ref{fig:3}~(c)), and together with a redshift of the resonant wavelength, $Q$-factor can be boosted from $\approx10^2$ to $10^4$ (see Fig.~\ref{fig:3}~(d)).

To clarify the reason for such a rapid wavelength shift of the resonances, we consider the modification of the FBZ and quasiguided modes dispersion with a change of the rotation angle. Figure~\ref{fig:3}~(e) schematically shows the dispersion of the moiré-induced modes (green, dashed lines) that pass through the $\Gamma$ point. A slight increase of $\alpha$ results in growth of moiré Bragg vector $G_{\mathrm{moire}}=\left|\mathbf{G}_{\mathrm{upper}}-\mathbf{G}_{\mathrm{lower}}\right|=\frac{4\pi}{a}\sin \frac{\alpha}{2}$,  deformation of FBZ and split of the modes (see Fig.~\ref{fig:3}~(f)). In order to bring them back to the $\Gamma$ point, we need to reduce the considered wavelength and increase the radius of the corresponding modes (see Fig.~\ref{fig:3}~(g)). Since the position of the FBZ boundary is rather sensitive to the angle, the energy of the modes becomes sensitive as well. Importantly, together with a movement of resonant modes corresponding Fourier harmonics shift in reciprocal space. In particular, if the harmonics ensuring coupling of the modes possess relatively large in-plane components, $\mathbf{k_\parallel}$, of the wavevector, they rapidly decay along the $z$-axis $\left(\propto e^{-\mathrm{Im}\sqrt{k^2-k_\parallel^2}z}\right)$. Consequently, the only channel of leakage through the scattering on the neighboring lattice is suppressed and quality factors of corresponding hybrid modes are rather high.

Finally, we theoretically study the near-field of the long-wavelength (low-energy) mode from Fig.~\ref{fig:3}. Figures~\ref{fig:4}~(a-c) demonstrate the distribution of the electric field $|\mathbf{E}|^2$ of the hybrid modes in stacks of different gap-size for the vertical section ($y=\mathrm{const}$) in which dielectric strips lie directly over the other strips. Figures~\ref{fig:4}~(d-f) consider the same modes for the vertical section in which strips of upper and lower sublattices alternate each other.
In both cases, the field is localized near the dielectric slabs with a notable enhancement in the vicinity of the edges of the dielectric strips. For the small value of the gap size ($g=20$~nm) we observe a significant concentration of the field between the sublattices, especially in Fig.~\ref{fig:4}~(d), which indicates strong coupling between their modes. Nevertheless, with a slight increase in the thickness, we immediately see that the field near each of the structures is determined primarily by their own eigenmodes, whereas the hybridization is responsible for their relative phases.
The symmetric intensity maps in Fig.~\ref{fig:4}~(a-c) might give a misleading impression that the considered moiré modes possess corresponding symmetry as well, but Fig.~\ref{fig:4}~(d-f) clearly shows that the modes evolve along the strips and none of the sections by itself can provide us with the full picture.


The distribution of the eigenmode's electric field intensity in five horizontal sections is illustrated in Fig.~\ref{fig:4}~(g-k). We consider the field in the middle of all three layers (see panels (h-j)),  and at a 50~nm distance from them (see panels (g) and (k)). The most important feature of the field is the standing wave clearly observed in panels (i) and (g,k). This pattern is associated with guided modes of both sublattices propagating in opposite directions of the $y$-axis. The field of standing waves in each specific section is slightly affected by the adjacent dielectric strips, which is clearly seen in Fig.~\ref{fig:4}~(g,i,k)). Interestingly, the larger the distance from the section to the lattice the clearer should be the picture of the standing waves due to the extinction of high-$k_\parallel$ Fourier harmonics. For the sections of upper and lowers sublattices (see Fig.~\ref{fig:4}~(h,j)), we observe rather complicated field distribution with plenty of peculiarities such as hot spots localized near the dielectric-vacuum interfaces.


In conclusion, we studied twisted stacks of identical photonic crystal slabs and demonstrated an emergence of specific moiré modes due to inter-layer interaction. These modes are very sensitive to the angle of sublayers' relative rotation, $\alpha$, and the gap size between them, $g$. In particular, we managed to tune the mode wavelength in the range of 300-600~nm and vary the quality factor from $10^2$ up to $10^5$ and this is not the limit. The demonstrated results show the moiré structures' potential for engineering high-quality optical resonances, their active tuning, and their application in optomechanical devices and chiral nanophotonics.


\textit{Acknowledgements.} This work was supported by the Russian Science Foundation (Grant No. 22-12-00351).



%

\end{document}